\documentclass{svproc}
\usepackage{url}

\begin{document}
\mainmatter              
\title{Fundamental Physics, the Swampland of Effective Field Theory and
Early Universe Cosmology}
\titlerunning{Swampland}  
%
\author{Robert Brandenberger}
\authorrunning{Brandenberger} 
%
\tocauthor{Robert Brandenberger}
\institute{Physics Department, McGill University, Montreal, QC, H3A 2T8, Canada\\
\email{rhb@physics.mcgill.ca}.
}

\maketitle              

\begin{abstract}

Cosmological inflation is not the only early universe scenario consistent with current
observational data. I will discuss the criteria for a successful early universe
cosmology, compare a couple of the proposed scenarios (inflation, bouncing
cosmologies, and the {\it emergent} scenario), focusing on how future
observational data will be able to distinguish between them. I will argue that
we need to go beyond effective field theory in order to understand the early
universe, and that principles of superstring theory will yield a nonsingular cosmology.


\end{abstract}
\section{Introduction}
In this talk I would like to convey three main messages. The first is that
the {\it inflationary scenario} is not the only early universe scenario which is
consistent with current observational data. The second message is that 
the inflationary scenario does not appear to naturally emerge from superstring
theory. On a positive note (and this is the third message), there are arguments
based on fundamental principles of superstring theory which indicate that the cosmology which
emerges from string theory will be nonsingular.

The two past decades have provided us with a wealth of data about the
structure of the universe on large scales. From the point of view of Standard
Big Bang cosmology most of the data cannot be explained. Why is the universe
close to homogeneous and isotropic on scales which at the time of recombination
had never been in causal contact? Why is the universe so close to being spatially
flat? These are the famous {\it horizon} and {\it flatness} problems of Standard Big
Bang cosmology. We now have detailed measurements of the small amplitude
inhomogeneities in the distribution of matter and radiation, most spectacularly the
high precision all sky maps of the cosmic microwave background (CMB) radiation \cite{Planck}.
The angular power spectrum of these anisotropies shows that the fluctuations
are scale-invariant on large scales and are characterized by {\it acoustic oscillations}
on smaller scales. What is the origin of these fluctuations?

The physics which yields the abovementioned acoustic oscillations in the angular
power spectrum of CMB fluctuations was discussed in two pioneering papers \cite{SZ,Peebles}.
These authors assumed the existence of a roughly scale-invariant spectrum of curvature fluctuations
on super-Hubble scales (the Hubble radius is $H^{-1}(t)$, where $H$ is the Hubble
expansion rate) at a time before recombination. These fluctuations are standing waves
which are frozen in until the time when the Hubble radius becomes larger than the
length scale of the fluctuations (in the linear regime fluctuations have constant wavelength
in comoving coordinates; hence, in the matter dominated epoch their physical wavelength
grows as $t^{2/3}$ while the Hubble radius grows at the faster rate $\sim t$). After they
enter they begin to oscillate. Modes which have performed an even (odd) number of half oscillations
by the time of recombination yield maxima (local minima) in the power spectrum. The
papers \cite{SZ,Peebles} date back to ten years before the development of inflationary cosmology. Both
the CMB acoustic oscillations and the {\it baryon acoustic oscillations} in
the power spectrum of matter fluctuations were predicted already then.

The question which was not addressed in \cite{SZ,Peebles} is the origin of the super-Hubble
fluctuations at early times. In Standard Big Bang cosmology the Hubble radius equals the
horizon, and hence having super-Hubble fluctuations appears to be acausal. Inflationary
cosmology \cite{inflation} was the first scenario to propose an origin \cite{ChibMukh} for these
fluctuations, but now we know that it is not the only one. In the following I will develop necessary criteria for
an early universe scenario to be able to explain the near homogeneity of the universe and
the origin of the observed cosmological perturbations. I will then introduce a couple of
early universe scenarios which satisfy the criteria. In Section 3 I will turn to the question of
which early universe scenario might emerge from superstring theory.
\section{Early Universe Scenarios}
The first criterion which a successful early universe scenario must satisfy is that the horizon (the
radius of the forward light cone of a point on the initial value surface) is much larger than the Hubble 
radius $H^{-1}(t)$ at late times. This is necessary to be able to address the horizon problem of
Standard Big Bang cosmology. In order to admit the possibility of a causal mechanism to generate
the primordial fluctuations, comoving scales which are probed with current cosmological observations
must originate inside the Hubble radius at early times. This is the second criterion. If the fluctuations
emerge as quantum vacuum perturbations (as they are postulated to in inflationary cosmology), then
scales we observe today must evolve for a long time on super-Hubble scales in order to obtain
the squeezing of the fluctuations which is necessary to obtain classical perturbations at late times (third
criterion). Finally (fourth criterion), the structure formation scenario must produce a nearly scale-invariant
spectrum of primordial perturbations (see e.g. \cite{RHBrev} for a more detailed discussion).

Inflationary cosmology \cite{inflation} is the first scenario which satisfies the four above criteria. During
the time interval $t_i < t < t_R$ during which the universe undergoes nearly exponential expansion, the
horizon expands exponentially while the Hubble radius remains almost unchanged. Since the physical
length of a fixed comoving scale also expands nearly exponentially during the period of inflation, scales 
which we observe today originate inside the Hubble radius as long as the period of inflation is sufficiently
long. Fluctuations are squeezed on super-Hubble scales for a long time, and the approximate time-translation
symmetry of the inflationary phase ensures that the spectrum of primordial fluctuations is
nearly scale-invariant \cite{Press,ChibMukh}.

Bouncing cosmologies provide a second scenario in which the four criteria for a successful early
universe scenario can be satisfied. In a bouncing scenario the horizon is infinite. The Hubble radius
decreases during the period of contraction and then increases during the period of expansion. As long
as the period of contraction is comparable in length to the period of Standard Big Bang expansion,
scales which we observed today emerge from inside the Hubble radius, thus allowing a possible causal
structure formation scenario. As in inflationary cosmology, there is a long period during which scales
propagate with super-Hubble length, thus enabling the squeezing of the fluctuations. There are (at least)
three classes of bouncing cosmologies. First, the {\it matter bounce} \cite{Fabio} in which there is a 
long phase of matter-dominated contraction. Second, there is the {\it Pre-Big-Bang} scenario \cite{GV}
in which contraction is driven by a field with an equation of state $w = 1$, where $w$ is the ratio of 
pressure to energy density. Finally, there is the {\it Ekpyrotic} scenario \cite{Ekp} in which contraction is obtained
by means of a scalar field with equation of state $w \gg 1$. There is a duality between the evolution of
curvature fluctuations in a matter-dominated phase of contraction and in an exponentially expanding
background \cite{Wands}. Hence, the matter bounce automatically leads to a roughly scale-invariant
spectrum of fluctuations. There is a duality in the evolution of scalar field fluctuations between an
cosmology with Ekpyrotic contraction and one of exponential expansion \cite{KOST2}. Hence, it is
also possible to obtain a scale-invariant spectrum of fluctuations. In the case of the Pre-Big-Bang
scenario it is possible to obtain a scale-invariant spectrum making use of axion fields \cite{axion}. See
\cite{Bounce-Review} for a detailed review of bouncing cosmologies. Ekpyrotic and Pre-Big-Bang
cosmologies produce a steep blue spectrum of primordial gravitational waves. Hence, on cosmological
scales the spectrum of primordial gravitational waves is predicted to be negligible. This contrasts
with the predictions of inflationary models which forecast a roughly scale-invariant spectrum.

A third scenario for early universe cosmology is the {\it emergent scenario} which is based on the assumption
that the universe emerged from an initial high density state in which matter was in global thermal
equilibrium. One toy model for this is {\it String Gas Cosmology} \cite{BV} in which it is assumed
that the universe loiters for a long time in a Hagedorn phase of a gas of fundamental strings, and there
is a phase transition to the expanding phase of Standard Big Bang cosmology (see e.g. \cite{SGrev}
for a review). In the emergent scenario, the horizon is infinite, and scales which are observed today
are trivially sub-Hubble in the emergent phase (since the Hubble radius is infinite in the limit
that the emergent phase is static). As discovered in \cite{NBV}, the spectrum of cosmological
perturbations originating from thermal fluctuations of the string gas is nearly scale-invariant. A
prediction with which String Gas Cosmology can be distinguished from simple inflationary models
is the tilt of the spectrum of primordial gravitational waves. Whereas inflationary models based on
a matter content which satisfies the usual energy conditions predict a slight red tilt of the spectrum,
String Gas Cosmology predicts a blue tilt $n_t$ satisfying a consistency relation $n_t = n_s - 1$,
where $n_s - 1$ is the tilt of the spectrum of curvature fluctuations \cite{BNPV}.

None of the early universe scenarios discussed above are without problems. In the case of
inflationary cosmology we can point to the {\it trans-Planckian problem} for fluctuations: if the
period of inflation is much longer than the minimal period which inflation has to last in order to
enable a causal generation mechanism of fluctuations, the length scale of all modes which are
currently observed today was smaller than the Planck length at the beginning of inflation
\cite{Jerome}. Thus, new physics must enter to give the initial conditions for the fluctuations.
As discussed in \cite{Yifu}, the matter bounce scenario is not a local attractor in initial condition
space: initial anisotropies blow up during the contracting phase. The Ekpyrotic scenario does
better in this respect: initial anisotropies decay and the homogeneous Ekpyrotic contracting
trajectory is a local attractor in initial condition space \cite{Erickson}. Note that in the case of 
{\it large field inflation}, the inflationary slow-roll trajectory is also a local attractor
\cite{Kung}. A key challenge for bouncing scenarios is that new physics is required to yield
the cosmological bounce. An important problem for the emergent scenario is to obtain dynamical
equations which describe the emergent phase.

Both inflationary and Ekpyrotic models are obtained in the context of Einstein gravity by
taking the dominant component of matter to be given by a {\it scalar field} $\varphi$ with
a potential $V(\varphi)$. To obtain slow-roll inflation the potential has to be very flat
\begin{equation} 
\frac{V'}{V} \, \ll \, m_{pl}^{-1} \, ,
\end{equation}
where the prime indicates a derivative with respect to $\varphi$, and where $m_{pl}$ is
the Planck mass. For models free of an initial condition fine tuning problem the field
$\varphi$ must roll over a field range $|\Delta \varphi| > m_{pl}$ during inflation. In contrast,
the Ekpyrotic scenario is based on scalar field matter with a negative and steep exponential 
potential. $V'/V$ is large in Planck units, and the scalar field rolls a distance smaller than $m_{pl}$.
I highlight this point in connection with the constraints on effective field theories involving
scalar fields which emerge from the considerations based on fundamental physics to be discussed in
the following section.

\section{Constraints from Fundamental Physics}

The evolution of the very early universe should be described by the best available theory which 
describes physics at the highest energies. There is evidence that all forces of nature might unify
at high energies. They must be described quantum mechanically. The best candidate for
such a quantum theory is {\it superstring theory}. Superstring theory is based on the assumption 
that the basic building blocks of nature are not point particle, but rather elementary strings.

The quantum theory of point particles is quantum field theory. There is a huge {\it landscape}
of quantum field theories: any number of space-time dimensions and fields is allowed, and any
shape of field potentials. Superstring theory is very restrictive. The number of space-time
dimensions is fixed, and the string interactions are constrained. At low energies, the physics
emerging from superstring theory should be describable by an {\it effective field theory}.
However, the set of effective field theories compatible with string theory is constrained by
what are known as the {\it swampland criteria}. Only theories consistent with these criteria
are admissible. The vast number of field theories are not - they are said to lie in the
{\it swampland} (see \cite{Palti} for a review). Note that at the moment these criteria are not
proven - they are educated guesses.

The first swampland criterion \cite{Vafa1} is that the field range over which a given
effective field theory is valid is constrained by $\Delta \varphi < {\cal{O}}(1) m_{pl}$. The
second condition \cite{Vafa2} is that, for a scalar field which is rolling and which dominates the
energy density of the universe, the potential cannot be too flat:
\begin{equation}
\frac{V'}{V} \, > \, c_1 m_{pl}^{-1} \, ,
\end{equation}
where $c_1$ is a constant of order one. This condition is opposite to what is required for
simple slow-roll inflation models. Hence, it appears that cosmological inflation is in tension
with superstring theory. A corollary of the second swampland condition is that a cosmological phase
dominated by a positive cosmological constant is not possible. Hence, Dark Energy cannot be
a cosmological constant \cite{Vafa3}. Scalar field models of Dark Energy are, however,
consistent with (but constrained by) the swampland conditions \cite{Vafa3,Lavinia}. 

In light of the tension between inflationary cosmology and the principles of string theory it
appears that we may need a new paradigm of early universe cosmology. Such a paradigm
should be based on the key new degrees of freedem and symmetries which differentiate
string theory from point particle theories. New degrees of freedom include the string oscillatory
and winding modes. Let us for simplicity consider the background space to be toroidal. Strings
on this space have momentum modes whose energies are quantized in units of $1/R$, where $R$
is the radius of the torus, winding modes whose energies are quantized in units of $R$, and
an tower of oscillatory modes whose energies are independent of $R$. Point particles only have
momentum modes. If we consider a box of strings in thermal equilibrium and compress the radius, 
then the temperature of the gas will initially increase since the energy of the momentum modes (which
are the light modes for large values of $R$) increases. Eventually it becomes thermodynamically
preferable to excite higher and higher energy oscillatory modes. The increase in temperature will
level off: there is a maximal temperature of a gas of strings, the {\it Hagedorn temperature} $T_H$
\cite{Hagedorn}. When $R$ decreases below the string scale, the energy will flow into the winding
modes (which are now the light modes), and the temperature will decrease. Hence \cite{BV}, 
thermodynamic reasoning indicates that there is no temperature singularity in a stringy early
universe cosmology.

String theory also features a new symmetry, {\it T-duality} symmetry. For a toroidal space, this implies
that there is a symmetry between a space of radius $R$ and a dual space of radius $1/R$ (in string
units) obtained by interchaning the momentum and winding quantum numbers. As already argued
in \cite{BV}, the number of position operators in a quantum theory of strings must be doubled
compared to a theory of point particles: there is one position operator which is the Fourier transform
of the momentum
\begin{equation}
|x> \, = \, \sum_n |p>_n \, ,
\end{equation}
where $|p>_n$ is the eigenstate of momentum which quantum number $n$ ($n$ ranging over the
integers), and a dual operator $|{\tilde{x}}>$ which is dual to the winding number eigenstates. Physical
length $l_p$ is measured in terms of $|x>$ if $R$ is large, but in terms of $|{\tilde{x}}>$ if $R$ is small. Hence,
as $R$ decreases from some large value towards zero, $l_p$ remains finite (it is an even function of 
${\rm{ln}}(R)$). This is another way to see the non-singularity of a stringy early universe cosmology.
 
The challenge for string cosmology remains to find consistent equations for the time-dependent
cosmological background. Einstein gravity is not applicable since it is not consistent with the T-duality
symmetry of string theory. In String Gas Cosmology \cite{BV} it was postulated that the universe emerges
from a quasi-static initial Hagedorn phase. Such a phase could emerge from a better understanding
of non-perturbative string theory. If we want to model the dynamics using an effective field theory, this
effective field theory must live in double the number of spatial dimensions as the topological background 
contains in order to take into account both the $|x>$ and $|{\tilde{x}}>$ coordinates. A candidate for such a
theory is {\it Double Field Theory} \cite{DFT}, a theory which is given by the action for a generalized
metric in doubled space. The cosmology which results if we couple the Double Field Theory action for
the background to ``string gas matter'' (matter which has an equation of state of radiative modes for 
large volumes of the $|x>$ space, and that of winding modes for a small volume) was recently analyzed
in \cite{us}. In this context it can be shown that the solutions in the string frame are non-singular.

\section{Conclusions and Discussion}

In the context of effective field theories of matter coupled to Einstein gravity, a number of
early universe scenarios have been proposed which can explain current observational data.
Inflationary cosmology is one of them, but not the only one. However, general considerations
based on superstring theory indicate a tension between fundamental physics and inflation. In
fact, they indicate that any approach based on effective field theory of matter coupled to
Einstein gravity will break down in the early universe, and that we need a radically
different approach which takes into account the new degrees of freedom and new
symmetries which distinguish string theories from point particles theories. I presented a toy
model which take these aspects into account which indicates that the cosmology emerging
from string theory will be non-singular, and that it may not include any phase of inflation.

\section{New Developments}

Since the date of the talk there has been a new development. Motivated by the
{\it trans-Planckian problem} \cite{Jerome} for fluctuations in inflationary cosmology
it has been suggested \cite{Bedroya} that fundamental physics may prohibit
trans-Planckian scales from ever exiting the Hubble horizon and becoming classical,
in analogy to Penrose's Cosmic Censorship Hypothesis \cite{Penrose}, which postulates that timelike
singularities must be hidden by horizons. As discussed in \cite{BBLV}, this 
{\it Trans-Planckian Censorship Conjecture} (TCC) severely constrains models of inflation. The
TCC sets an upper bound on the duration of inflation while, in order for inflation to be able to
explain the formation of structure on the largest scales observed today, the period of inflation
has to be sufficiently long. The two constraints are consistent only if the scale of inflation is
less than about $10^9 {\rm GeV}$. This, leads to a very small tensor to scalar ratio
$r < 10^{-30}$, implying that if a stochastic background of gravitational waves is
discovered on cosmological scales, it cannot be due to inflation. Ways to alleviate
this tension by adding new dials to inflation have been suggested in \cite{new}.

%
%

\end{document}